\documentclass{article}
\usepackage{emulateapj}
\usepackage{float,epsfig}

\newcommand{\re}{$r_e$}
\newcommand{\LA}{\mbox{\raisebox{-0.6ex}{$\stackrel{\textstyle<}{\sim}$}}}
\newcommand{\GA}{\mbox{\raisebox{-0.6ex}{$\stackrel{\textstyle>}{\sim}$}}}
\newcommand{\cxo}{{\sl Chandra}}
\newcommand{\xmm}{{\sl XMM/Newton}}
\newcommand{\ngc}{{NGC~2403}}
\newcommand{\msun}{$M_{\odot}$}
\newcommand{\ergl}{erg~s$^{-1}$}
\newcommand{\hi}{H{\sc i}}
\newcommand{\hii}{H{\sc ii}}
\newcommand{\ha}{H$\alpha$}

\newcommand{\hst}{{\sl Hubble}}
\newcommand{\ros}{{\sl ROSAT}}
\newcommand{\cxou}{CXOU~J073650.0+653603}
\newcommand{\etal}{et al.}
\slugcomment{Submitted to Astrophysical Journal}

\begin{document}

\title{Discovery of a Transient X-ray Source in the Compact Stellar Nucleus of 
NGC 2403}

\author{
Mihoko~Yukita\altaffilmark{1},
Douglas~A.~Swartz\altaffilmark{2},
Roberto~Soria\altaffilmark{3}, and
Allyn~F.~Tennant\altaffilmark{4}
}
\altaffiltext{1}{University of Alabama in Huntsville, Dept. of Physics,
    Huntsville, AL, USA}
\altaffiltext{2}{Universities Space Research Association,
    NASA Marshall Space Flight Center, VP62, Huntsville, AL, USA}
\altaffiltext{3}{Harvard-Smithsonian Center for Astrophysics,  
    Cambridge, MA, USA}
\altaffiltext{4}{Space Science Department,
    NASA Marshall Space Flight Center, VP62, Huntsville, AL, USA}

\begin{abstract}
We report the discovery of an X-ray source coincident with the
 nuclear star cluster at the dynamical center of 
 the nearby late-type spiral galaxy \ngc.
The X-ray luminosity of this source varies from below detection
 levels, $\sim$10$^{35}$~\ergl\ in the $0.5-8.0$~keV band, to
 7$\times$10$^{38}$~\ergl\ on timescales between observations 
 of $<$2~months.
The X-ray spectrum is well-fit by an accretion disk model 
 consisting of multiple blackbody components and corresponding physically
 to a compact object mass of \GA 5~\msun.
No pulsations nor aperiodic behavior is evident in its X-ray light curve
 on the short timescales of the individual observations.
The X-ray properties of the source are more similar to those of 
 the nuclear source X-8 in M33, believed to be a low-mass X-ray binary,
 then to those of the low-luminosity active galactic nucleus in NGC~4395.
The brightness of the nuclear star cluster, $M_I \sim -11.8$~mag, is typical of 
 clusters in late-type spirals but its effective radius, $r_e \sim 12$~pc,
 is several times larger than average indicating a relatively relaxed 
 cluster and a low probability of a central massive object.
The cluster has a mass \GA10$^{6.5}$~\msun\ and an
 age of $\sim$1.4~Gyr estimating from its observed colors and brightness. 
\end{abstract}


\section{Introduction}

Tight correlations have been measured between 
 the masses of central supermassive black holes in early-type galaxies 
 and their bulge mass (Magorrian \etal\ 1998; H\"{a}ring \& Rix 2004),
 luminosity (Kormendy \& Richstone 1995),
 central velocity dispersion (Ferrarese \& Merritt 2000; Gebhardt \etal\ 2000;
  Tremaine \etal\ 2002),
 and central light concentration (Graham \etal\ 2001).
It is unclear if these correlations extend to less massive galaxies and 
 to galaxies of later morphological type.
If so, do the scaling properties still hold? That is, do disk-dominated 
 late-type spirals with small bulges host intermediate-mass black holes 
 or is there a minimum galaxy mass below which  
 black holes fail to form and/or to grow?

High-resolution \hst\ images (Carollo \etal\ 1998; B\"{o}ker \etal\ 2002, 2004)
 show that the majority, $\sim$75\%, of late-type spirals have  distinct compact
 stellar nuclei or nuclear star clusters (NSCs). 
These are not, however, a smooth extension from the massive bulges
 of early-type galaxies.
They are much more dense and more compact (Walcher \etal\ 2005);
 hence, while typical NSC masses can be comparable to those of
 dwarf spheroid galaxies, they are 4 orders of magnitude denser. 
In this sense, NSCs are more like  
 Milky Way globular clusters with scaled-up masses and densities. 
In addition, they are much more luminous than the old globular clusters
 (B\"{o}ker \etal\ 2004) due in large part to the comparatively young age, 
 100~Myr, of their most luminous stellar component (Rossa \etal\ 2006).
Still, their dynamical masses 
 suggest a hidden older population is present as well (Walcher \etal\ 2005;
 Rossa \etal\ 2006) so that they may grow through occasional bursts of 
 star formation and may be
 forming stars in the current epoch at low levels.

While the NSCs are, by definition, at or very near the dynamical centers
 of their host galaxies (e.g. B\"{o}ker \etal\ 2002), it is not certain 
 if they originated there nor whether or not they are presently 
 accreting gas and forming stars. 
It has been suggested that 
 some of the most massive globular clusters may be relic dE nuclei tidally 
stripped and absorbed into larger galaxies (e.g. Freeman 1993;
 Layden \& Sarajendini 2000).
If they commonly host (intermediate-mass)  black holes, then
 they may be potential candidates for ultraluminous X-ray sources
 (King \& Dehnen 2005) and, if they migrate to the cores of the larger
 host galaxies, may be the seed (intermediate-mass) black holes that evolve
 into supermassive black holes.

The SAB(s)cd galaxy
\ngc\ is a member of the M81 group of galaxies at a distance 
 $D=3.2$~Mpc (1\arcmin$=$1~kpc; Madore \& Freedman 1991).
 \ngc\ lacks a central bulge 
 but does host a luminous compact NSC
 (Davidge \& Courteau 2002).
We derive physical properties for this cluster from recent \hst\
 Advanced Camera for Surveys observations and other archival data in
 \S~\ref{s:cluster}. 
We have discovered a bright transient X-ray source within the
 \ngc\ NSC in archival \cxo\ and \xmm\ observations. 
The nuclear transient is easily resolved from nearby X-ray sources 
 including the well-known ultraluminous X-ray source located about 
 2.\arcmin 6 to the west of the nucleus (e.g., Swartz \etal\ 2004).
We present the X-ray spectra and light curve of the nuclear source 
 in \S~\ref{s:Xtrans}.
The optical properties of the NSC and the X-ray properties 
 of the transient source are compared to other nearby compact nuclei 
 in \S~\ref{s:compare}.
We find the NSC is older and less compact than typical for late-type galaxies.
The properties of the X-ray source are consistent with 
 an X-ray binary containing an $\sim$5~\msun\ compact object or larger
 accreting from a low-mass companion. 
Further discussion is given in \S~\ref{s:discuss}

\section{The Nuclear Star Cluster in \ngc} \label{s:cluster}

The contrast between the nucleus of \ngc\ and the underlying 
 galaxy disk is clearly visible in optical and near-IR images where
 it appears extended at high resolution (Figure~\ref{f:opt_image}).
We used the method described in B\"{o}ker \etal\ (2004) to analyze the morphology  
 of the cluster.

Calibrated \hst\ images acquired with the Advanced Camera for Surveys (ACS)
 operated in WFC mode were obtained from the Multimission Archive at 
 STScI\footnote{http://archive.stsci.edu/}.
These images were already cleaned of cosmic rays and bad pixels and were   
 corrected for geometric distortion; 
 we performed no further processing of the data.
The data were taken 2004-08-17 and include images through the F475W, F606W, 
 F658N, and F814W filters (dataset identifiers J90ZX1010 through 1040).

\begin{center}
\includegraphics[angle=0,width=\columnwidth]{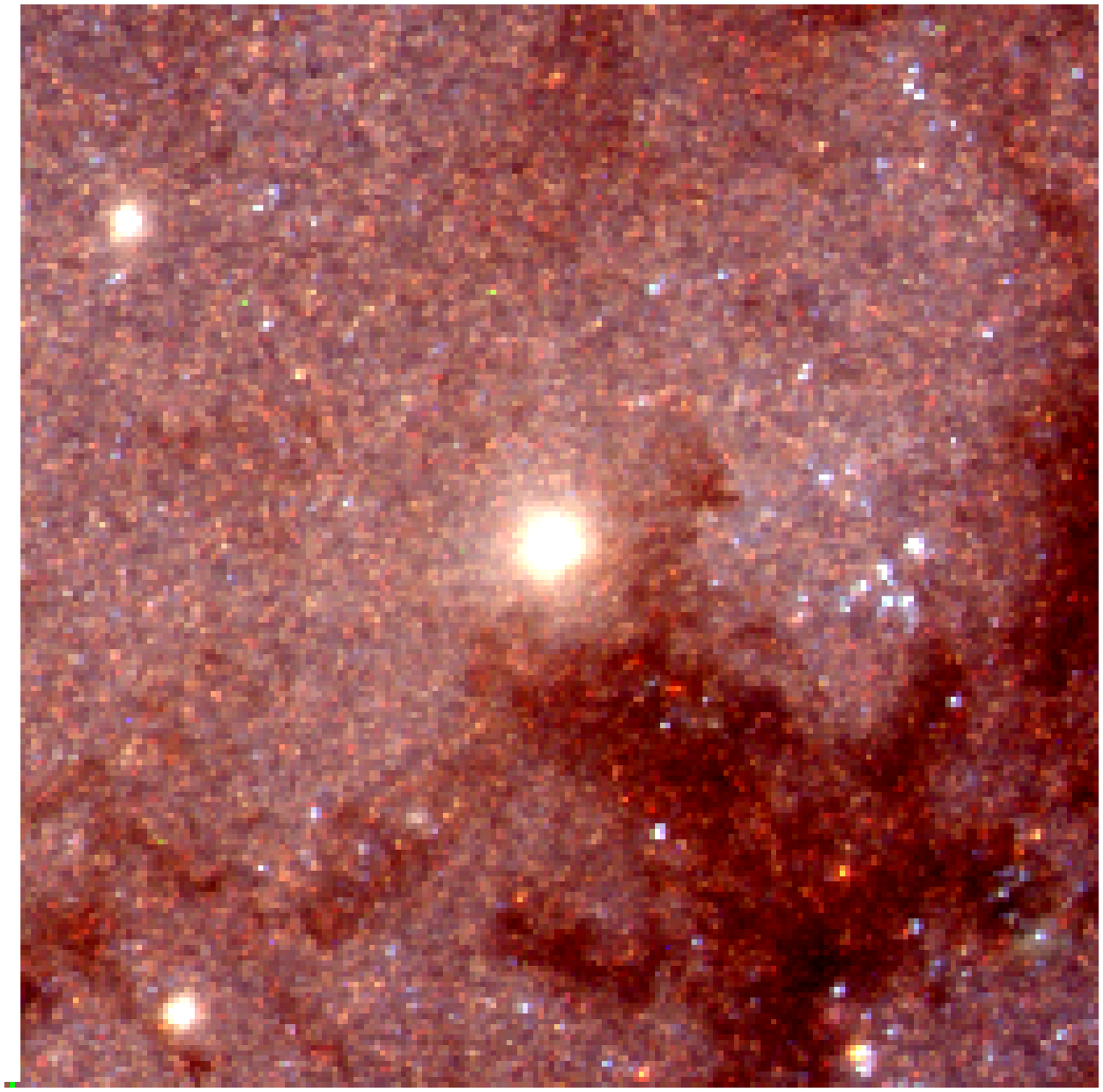}
\figcaption{3-color combined \hst/ACS image of the 20\arcsec$\times$20\arcsec\ region centered on the nuclear star cluster in \ngc. Colors blue, green, and red correspond to 
the filters F475W, F606W and  F814W, respectively. The image has been
 rotated for display purposes such that north is up and east is to the left.
\label{f:opt_image} }
\end{center}

The 200\arcsec$\times$200\arcsec\ \hst/ACS field contained both 
 SN~2004dj (the target of the original investigation) and the nucleus of \ngc.
Although the location of SN~2004dj is known to high precision
(Beswick \etal\ 2005) and the ACS resolution is $\sim$0.\arcsec 1 (with 0.\arcsec 05~pixel$^{-1}$ sampling), the supernova is saturated in these
 images and this is the dominant contribution to the uncertainty in the 
 position of the nucleus:
Our best estimated position for the nucleus is 
R.A.$=$7$^{\rm h}$36$^{\rm m}$50.$^{\rm s}$070, 
Decl.$=+$65$^{\circ}$36$^{\prime}$3.$^{\prime\prime}$54 (J2000.0)
 with an uncertainty of $\sim$0.\arcsec 1 radius 
 based on a circular Gaussian model fit to the location of the supernova
 in the F606W filter image.
The position of the nucleus is consistent with the kinematic center of the galaxy  
determined by Fraternali \etal\ (2002) from the \hi\ rotation curve.

A model point spread function (PSF) of the \hst/ASC detector at the location of the source was constructed using TinyTim (Krist \& Hook 1997) assuming a spectral energy distribution equivalent to that of an A5V star (following B\"{o}ker \etal\ 2004). We then fit analytic models to the surface brightness distribution using the program ISHAPE (Larsen 1999) which convolves the model PSF with these analytic profiles. 
ISHAPE\footnote{ISHAPE documentation is available from http://www.astro.uu.nl/$\sim$larsen/baolab/} 
 allows for circular or elliptical models but cannot reproduce the 
 azimuthal asymmetries clearly seen in the \hst\ images of \ngc.
Therefore, we restricted our selection to circular 
Moffat and King model profiles.
The standard power indices and concentration parameters for these models
 (see Larsen 1999) provided poor fits to the data.
A trial King model applied to the azimuthally-averaged radial profile
 suggested a Moffat model with power index of 1.35 would improve the model
fit. This did give the best fit of all our trials.
Figure~\ref{f:ishape} displays this model, the data, and the fit residuals.
Following B\"{o}ker \etal\ (2004), we derive an effective 
 radius of $r_e = 11.7\pm0.1$~pc from the ISHAPE fitting.
This is larger than the average, $\sim$3.5~pc, found by 
 B\"{o}ker \etal\ (2004) for a sample of 39 NSCs in late-type galaxies.


We estimate the intrinsic, background-subtracted, $I$-band luminosity to be
 1.1$\times$10$^{39}$~\ergl\ in a $2\pi r_e^2$ region centered on the NSC 
 after correcting for an extinction of $E(B-V)=0.2$~mag (see below).  
The average surface brightness, 
 $I_e=323$~$L_{\odot}$~erg~s$^{-1}$~pc$^{-2}$, is near the low luminosity 
 range of the NSCs 
 in Sd galaxies in the B\"{o}ker \etal\ (2004) sample and near the high 
 luminosity end of Milky Way globular clusters (cf. their Fig.~5). 
The observed absolute blue magnitude is M$_{B} = -9.3$~mag
 (giving an extinction-corrected M$_B^{\rm o} = -10.2$~mag, see below).

\begin{center}
\includegraphics[angle=0,width=\columnwidth]{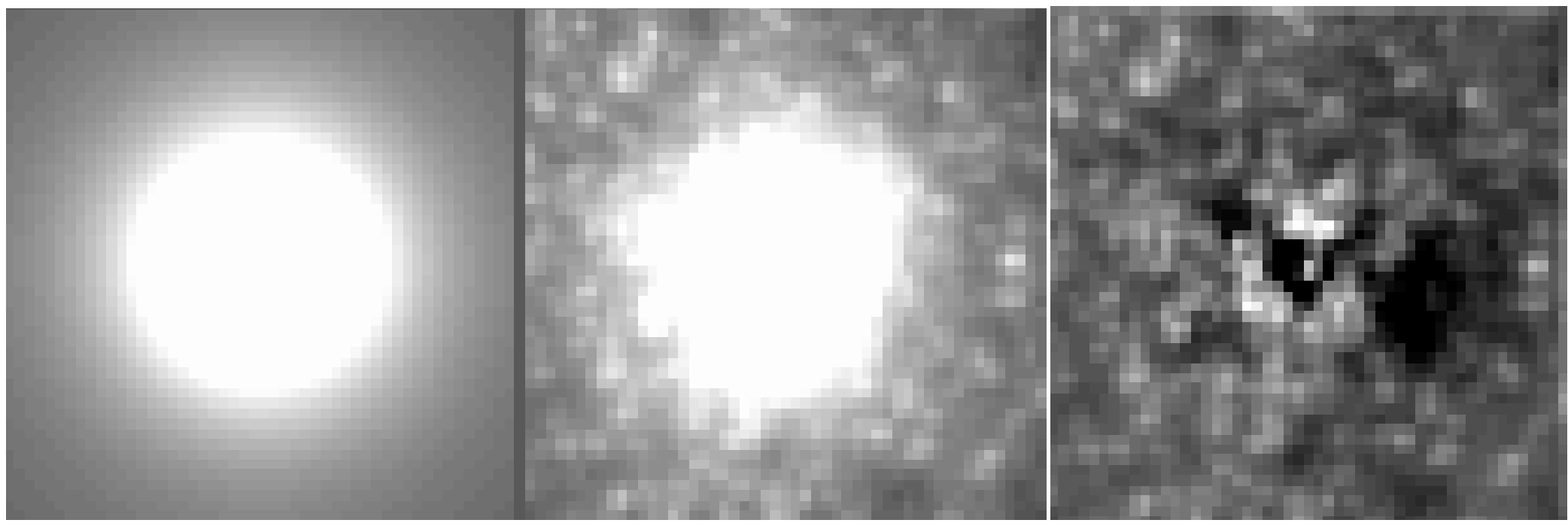}
\figcaption{ISHAPE (Larsen 1999) model fit to the $I$-band surface brightness distribution of the nuclear star cluster in \ngc. {\sl From left to right}: Model, data, and fit residuals. Grayscale of model and data cover the same range. The model is a Moffat profile with a power index of 1.35 (see text).
\label{f:ishape}}
\end{center}

Lacking an optical spectrum of the \ngc\ nucleus
 we cannot accurately constrain the mass and age of the cluster.  
However, we can obtain estimates of the cluster age from the observed 
 colors and the cluster mass from M$_B^{\rm o}$
 using the stellar synthesis models of, e.g., Leitherer \etal\ (1999).
The observed colors are (using the conversions from \hst/ACS to 
 Johnson-Cousins given by Sirianni \etal\ 2005)
 $B-V = 1.0$~mag, $V-I = 1.2$~mag, 
 and $J-K = 0.6$~mag (the latter from Davidge \& Courteau 2002).
These colors can be self-consistently reproduced by a single starburst 
 episode aged $\sim$1.4~Gyr viewed through an extinction of $E(B-V)= 0.2$~mag.
They cannot be reproduced by models with continuous star formation nor can 
 the metallicity of the cluster be radically different from the solar value.
The implied intrinsic colors are consistent with those of G2V to K0V stars.
As shown in \S~\ref{s:x_spectrum}, the extinction estimate is consistent with 
 the hydrogen column density towards the NSC X-ray source according to 
 the best-fitting X-ray spectral model. 

Rossa \etal\ (2006) show there is a strong correlation between the observed
 $B-V$ color of NSCs and their luminosity-weighted cluster age, $\tau_{\rm L}$.
Using their linear correlation coefficients gives
 $\tau_{\rm L} = 10^{9.8 \pm 0.4}$~yr. 
The age-mass correlation (Rossa \etal\ 2006) then gives 
 $\log M = 10^{8.3\pm1.8}$~\msun\ for the total mass of the NSC. 
These values are much higher than expected for a late-type spiral galaxy
 like \ngc\
 based on the (weaker) correlations Rossa \etal\ (2006) find 
 between cluster mass or cluster age and galaxy morphological type.

There is no correction for extinction in the work of Rossa \etal\ (2006).
One possibility is that the NSC in \ngc\ is more reddened than typical.
Figure~\ref{f:opt_image} shows the cluster is, in fact, at the edge of a dark 
 region however our best-fitting extinction is only $E(B-V)= 0.2$~mag.
Using the implied intrinsic $B-V = 0.8$~mag color reduces the estimated age to
 $\tau_{\rm L} = 10^{9.3 \pm 0.3}$~yr and the cluster mass to 
 $M = 10^{7.6\pm1.7}$~\msun. 
This age is consistent with the age estimated above from the colors using
 the starburst models of Leitherer \etal\ (1999). 
A mass estimate can also be deduced from the age and intrinsic blue 
 magnitude using these starburst models. 
The mass estimated in this way is $M \sim 10^{6.4}$~\msun\
 depending weakly on starburst metallicity and IMF.

The age estimated here is much older than the $\sim$100~Myr age deduced by
 Davidge \& Courteau (2002) for individual AGB stars in the vicinity
 of the nucleus. 
(Many of these stars are visible as blue sources in the composite \hst\ image
 of Figure~1.)
Davidge \& Courteau (2002) point out that the $J-K$ color of the cluster
 is somewhat bluer than these surroundings suggesting an age gradient.
According to Leitherer \etal\ (1999), the $J-K$ color for a starburst
 becomes bluer with age (after an early reddening phase)
 implying that the NSC in \ngc\ is older than its surroundings.
This is consistent with our calculations.

Although the bulk of the star formation in the cluster appears to have 
 occurred some $\sim$1.4~Gyr ago, 
 there may be a low level of current star formation present
 (and, likely, an underlying older population of stars as well).
For star-forming clusters, the 
 current star formation rate can be estimated from the \ha\ luminosity. 
We have constructed a continuum-subtracted \ha\ image from the \hst/ACS
 images and find no net \ha\ emission; 
 the 2$\sigma$ upper limit to the \ha\ luminosity 
 within a 1.\arcsec 5 radius circle about the NSC is 3$\times$10$^{35}$~\ergl. 
Thus there is no evidence for current star formation in the NSC.

\section{The Transient X-ray Source in the Nucleus of \ngc } \label{s:Xtrans}

The nuclear region of 
 \ngc\ was observed with \cxo\ ACIS-S four times over a 3.7 year interval
 and three times with \xmm\ within this same time interval. 
Table~1 provides a log of these observations.
Earlier observations with the {\sl Einstein}, {\sl ASCA}, and \ros\ 
 Observatories show no evidence for a nuclear X-ray source although the 
 quality of the data is poor compared to the recent \cxo\ and \xmm\ 
 observations.

We obtained level 1 event lists from the \cxo\ data 
archive\footnote{http://cxc.harvard.edu/cda/} and reprocessed them using the Chandra Interactive Analysis of Observations (CIAO) version 3.3.0.1 tool {\tt acis\_process\_events} and the calibration database (CALDB) version  3.2.1. Reprocessing removed pixel randomization and applied CTI and 
time-dependent-gain corrections.  We then filtered the data of events with   non-{\sl ASCA} grades and bad status bits as well as hot pixels and columns.    

The European Photon Imaging Camera (EPIC) event lists from the \xmm\ observations were obtained from the HEASARC data 
archive\footnote{http://heasarc.gsfc.nasa.gov/db-perl/W3Browse/w3browse.pl}.  
We processed the event lists using the current calibration files with the routines in SAS 6.5.0.  

We checked for periods of high background that could affect our spectra and timing analysis; excluding some short intervals from the \xmm\ datasets.  The final Good Time Intervals for data used in our analysis are listed in Table~1.

\subsection{X-ray Light Curve}

A bright, point-like X-ray source was detected in several of the observations
 at the location of the nuclear star cluster. 
We used the common source, SN~2004dj, to match the location of the X-ray
 source and the NSC. The X-ray source is well within the $>$1\arcsec\
 optical extent of the cluster.
Figure~\ref{f:X_lc} shows the 0.5~--~8.0~keV light curve of the source,
 designated \cxou, constructed from the average flux during 
 each individual observation based on spectral fits (\S~\ref{s:x_spectrum}). 
For observations in which the source was not detected, 
 a 2$\sigma$ flux upper limit was calculated by scaling from the fitted
 measurements by the background-subtracted count rate in an appropriate
 source region.
The corresponding observed X-ray luminosities, assuming a distance of 3.2~Mpc,
 are tabulated in Table~1.
The X-ray source was undetected in five observations and orders of magnitude above (\cxo) detection thresholds in three observations which qualifies 
it as a transient source (e.g. van~Paradijs \& McClintock 1995).
The shortest measured interval between a detection and a non-detection is
 21~days (observations number 5 and 6 of Table 1) but the \xmm\ upper limit
 is not particularly compelling. The time between the \cxo\ non-detection 
 observation number 4 and the detection observation 6 is 41~days.

We also inspected the light curve from observations in which the source 
 was detected for variability during the observation. 
These X-ray light curves, binned into 1~ks intervals, displayed no 
 conspicuous variability and were formally consistent with a constant flux 
 model.
Searches for short-term aperiodic variability using 
 a power-spectrum analysis found no excess power above the Poisson noise.
Searches for coherent pulsations, using the $Z_n^2$ statistics (Buccheri 
 \etal\ 1983), detected nothing significant.
 
\begin{center}
\includegraphics[angle=-90,width=\columnwidth]{f3.eps}
\vspace{10pt}
\figcaption{Lightcurve of the X-ray source in the nucleus of \ngc.
Time is measured since the first \cxo\ observation on 2001-04-17. 
Upper limits for non-detections are represented with arrows. 
Crosses denote \cxo\ observations.
Triangles denote \xmm\ observations. Error bars denote two-sided errors.
Table~1 provides the numerical values
shown here.
\label{f:X_lc}}
\end{center}

\subsection{X-ray Spectrum} \label{s:x_spectrum}

Models were fit to observed spectra of the source.
Table~2 lists the models attempted, the resulting best-fit model parameters
 with 90\% confidence extremes for a single interesting parameter,
 and the fit statistic for each of the three observations in which the source
 was detected.
\cxou\ was most luminous during the \xmm\ observation of 2003 September.  
The spectrum from this observation is shown in Figure~\ref{f:X_xmmSpec}.
Following the procedures outlined 
 in Page \etal\ (2003),
the average \xmm\ spectrum from the three detectors was modeled in XSPEC
 using the $\chi^2$ fit statistic. 
Absorbed {\tt diskbb} and {\tt powerlaw} models were applied.
Adding additional model components 
 did not significantly improve the fit statistics according to the F-test.
We used an unbinned spectrum with C-statistic in XSPEC in fitting 
 the low-count \cxo\ observations. 
Only single-component models were applied to the \cxo\ spectra.
It was necessary to fix the 
 multiplicative absorption model component at the 
 Galactic value $n_H = 4 \times 10^{20}$~cm$^{-2}$ in fits to some of the \cxo\
 data to prevent convergence to an unphysically low value.

The best-fitting model in all cases is the {\tt diskbb} model
 (Makishima \etal\ 1986, 2000) representing a spectrum dominated by a 
 geometrically-thin accretion disk which extends to the innermost 
 stable orbit often referred to as a high/soft state 
 (Remillard \& McClintock 2006).  
The model fit parameters are proportional to the inner
 disk temperature and radius or, alternatively, 
 can be expressed in terms of the 
 mass of the central object. 
These parameters are listed in Table~2. They suggest a compact 
object mass, $M \sim 2.3/T_o^2 (L_{\rm disk,bol}/10^{38})^{1/2}$ 
 where $T_o \sim T_{\rm in}$ is the disk color temperature, 
 of about 5~\msun\ and that the peak observed luminosity is
 therefore about the Eddington value. 
The mass could be up to about a factor-of-two larger depending on viewing 
 angle, black hole spin, and hardening factor.

\begin{center}
\includegraphics[angle=-90,width=\columnwidth]{f4.eps}
\vspace{10pt}
\figcaption{\xmm\ spectrum of \cxou\ with the best fit disk-blackbody model
({\sl upper panel}) and fit residuals.
 \label{f:X_xmmSpec}}
\end{center}

\section{Comparison to Other Nuclear Star Clusters} \label{s:compare}

As mentioned in the Introduction, some 75\% of late-type spirals host 
 NSCs. 
In this section we briefly compare \ngc\ to two well-studied objects
 with very different interpretations for their X-ray emission mechanisms:
NGC~598 (M33), 
 another Scd galaxy with a NSC and a bright point-like nuclear X-ray source,
 is thought to contain a luminous X-ray binary; and
NGC~4395, the optically least-luminous broad-line AGN known,
 is thought to contain a ``proper'' nuclear black hole at the low end of 
 their mass distribution ($M \sim 10^5$~\msun). 

\subsection{M33 X8}

NGC~598 is a disk-dominated late type spiral galaxy as is \ngc.
Its NSC is as luminous ($M_B=-10.2$, Kormendy \& McClure 1993) as that of \ngc\ 
 but much more compact (FWHM $<$1 pc, Gordon \etal\ 1999).
Analysis of the cluster optical spectrum shows it 
 formed from two major starburst episodes; one $\sim$1~Gyr ago and  
 the other occurring only 40$-$70~Myr in the past (O'Connell 1983; 
 Gordon \etal\ 1999; Long \etal\ 2002).
The best estimated upper limit for the mass of any central compact object 
 in the cluster is $M_{\rm BH}$$\sim$1500~\msun\ based on modeling the narrow-slit
 \hst\ spectrum (Gebhardt \etal\ 2001).

The nucleus contains a bright unresolved X-ray source,
 designated X-8 (Trinchieri \etal\ 1988), with a luminosity as high as
 $\sim$2$\times$10$^{39}$~\ergl\ in the 0.5$-$10~keV band
 (Foschini \etal\ 2004).
It has long been debated whether this source could be a weak AGN.
The optical spectrum shows no evidence of AGN activity although a
 strong component would be expected if the measured X-ray flux is extrapolated 
 using the known X-ray to optical flux relationship for AGNs 
 (Long \etal\ 2002).
The current consensus is that X-8 is a luminous 
 X-ray binary in the NSC of NGC~598
 rather than a weak AGN (e.g., La~Parola \etal\ 2003).
The mass of the compact object must be $\sim$10~\msun\ or more unless it is
 radiating above the Eddington limit in the X-ray band.

The source X-ray flux has remained at about its current level since discovery 
 although it does vary by about a factor of two on short timescales.
This is in contrast with many low-mass XRBs which are soft X-ray transients that
 vary by factors of 100 or more on timescales of weeks or months (as does the
 nuclear source in \ngc).
It has been suggested that M33 X-8 is similar to GRS~1915$+$105 
 (e.g., Long \etal\ 2002; Dubus \& Rutledge 2002)
 in that both are steady sources with comparable X-ray luminosities
 and spectral states.
Both are also radio sources (Dubus \& Rutledge 2002).
GRS~1915$+$105 is a low-mass X-ray binary with a 14$\pm$4~\msun\ compact 
 object accreting from a low mass, evolved (K or M giant) companion
 (Greiner, Cuby, \& McCaughrean 2001).

\subsection{NGC 4395}

NGC~4395 is the optically least-luminous AGN known.
It is a SA(s)m dwarf galaxy with a distinct though weak type 1 Seyfert spectrum
 (Filippenko \& Sargent 1989).
The nuclear brightness is only $M_B \sim -11$~mag which makes it only
 twice as bright as the NSC in \ngc.
Filippenko \& Ho (2003) estimate the mass of its compact object to be less than
 $\sim$10$^5$~\msun.
Its X-ray spectrum is a moderately absorbed, 
 $N_H$$=$(1.2$-$2.3)$\times$10$^{22}$~cm$^{-2}$, hard power law 
  ($\Gamma$ \LA 1.5, Shih \etal\ 2003; $\Gamma\sim0.6$, Moran \etal\ 2005).
The absorption-corrected 2$-$10~keV luminosity is 
 nearly 10$^{40}$~\ergl\ (Iwasawa \etal\ 2000; Moran \etal\ 2005).
It is highly variable in X-rays (Moran \etal\ 1999, 2005; Shih \etal\ 2003) exhibiting 
 factor of 10 changes in X-ray flux in less than 2~ks.

The low power law index, moderate absorption, and rapid flux variability
 of the AGN in NGC~4395 is unlike the behavior of the nuclear source in \ngc\
 in all respects.

\section{Discussion} \label{s:discuss}

We have presented a new study of the NSC in \ngc.
We discovered that it contains a transient X-ray source;
its X-ray luminosity varies from $\sim$7$\times$10$^{38}$ to 
 \LA 5$\times$10$^{34}$ \ergl\
over a few months. The question is whether this accreting source
is the galaxy's nuclear black hole or, instead, a lower-mass
black hole formed in the star cluster from stellar processes.

Its luminosity is consistent both with a stellar-mass black hole (BH)
and with a supermassive black hole (SMBH) emitting a few orders of magnitude
below its Eddington limit (as are the large majority
of SMBHs in the Local Universe). Its X-ray variability
pattern, over at least 4 orders of magnitude over
a few months to years, is more consistent with typical
X-ray binaries. Its X-ray spectrum when the source is bright
 is exactly what we expect from
a BH of mass $\sim$5$-$10~\msun\ in the thermal dominant
state ($L_{\rm X} \sim 8\times10^{38}$~\ergl\ $ \sim L_{\rm Edd}$, 
$T_{\rm in}\sim 1$~keV,
$R_{\rm in}(\cos \theta)^{1/2} \sim 50$~km). SMBHs may have comparable
X-ray luminosities but tend to have power-law spectra
in the X-ray band, because their characteristic disk
temperatures are \LA 0.1 keV. We conclude that the X-ray
source in the NSC of \ngc\ is more likely to be
an ordinary X-ray binary, similar to the nuclear
source in the NSC of M33.

X-ray studies of nuclear sources associated with NSCs
provide an additional tool to understand and quantify
the relation between the formation of galactic bulges
and the presence and properties of a central ``compact
massive object'' (CMO). Ideally, one should look for correlations
that extend unbroken from large ellipticals to small
late-type disk galaxies. One such fundamental
correlation seems to hold between the bulge mass
and the CMO mass (Ferrarese \etal\ 2006; Wehner \& Harris 2006), where
the CMO is a supermassive BH in ellipticals
and early-type spirals and a NSC in late-type spirals.
The CMO mass is $\sim$1/500 of the spheroidal mass
(bulge mass in spirals). Qualitatively, this suggests
that the initial galaxy assembly processes led to
the rapid collapse of a nuclear gas component into a BH
in the more massive galaxies, while a star cluster
was formed in galaxies with a shallower gravitational
potential, where the central accumulation of gas
occurred more slowly.

However, many questions remain open. Some relatively
massive ellipticals do have NSCs (Rossa \etal\ 2006;
although they have not been found in galaxies brighter
than $M_B \sim -20$~mag; Ferrarese \etal\ 2006).
The range in which NSCs and SMBHs overlap may span
several orders of magnitude: a SMBH with $M_{\rm BH} \sim 3\times 10^6$~\msun\
is the CMO in the center of the Milky Way, and a SMBH
with a mass $\sim 10^5$~\msun\ is present in NGC~4395, but
NSCs as massive as a few $10^8$~\msun\ have been
found in some elliptical galaxies (Rossa \etal\ 2006).
It is not known whether the same galaxy can harbor
both an SMBH and a NSC, or if they are mutually
exclusive. It is also not yet clear which of
the two classes of objects defines the more fundamental
correlation with the galactic bulge; it was suggested
(Rossa \etal\ 2006) that NSCs in large galaxies
may be a factor of 3 more massive than SMBHs,
for a given bulge mass. Furthermore, it is not known
whether a linear relation between bulge and CMO
masses may extend all the way down to bulgeless
Scd galaxies (implying the existence of CMOs
with masses as low as $10^3-10^5$~\msun), or if
the relation breaks down for CMO masses \LA $10^6$~\msun.

From its optical brightness and colors, we estimate
a stellar mass of $\sim$2$\times$10$^6$~\msun\ for the NSC in \ngc,
comparable to the mass of the SMBH in the Milky Way,
even though \ngc\ is bulgeless. Although our
X-ray study suggests that the nuclear X-ray source
is not from a SMBH, this does not rule out the presence
of a SMBH inside the NSC: it may simply be quiescent.
In that case, it would be among the faintest SMBHs,
with a luminosity \LA 10$^{-10}$~$L_{\rm Edd}$.
In a work currently in preparation, we shall
use an X-ray survey of galactic nuclei to determine
whether galaxies with NSCs may also have signs
of AGN activity.

Even if we accept that a late-type galaxy such as \ngc\
formed a NSC but no SMBH, there is a strong possibility
that the NSC contains stellar-mass BHs formed via
ordinary stellar processes, or perhaps an intermediate-mass BH (IMBH)
formed via runaway core collapse and stellar coalescence
(if the NSC was sufficiently compact). Then, those BHs
could have grown over a Hubble time, via accretion
during large-scale gas inflows towards the galactic center,
via coalescence of stellar mass BHs inside the NSC,
or via orbital decay and coalescence of accreted
satellite galaxies or primordial halo remnants.
Processes involving BH mergers may lead to the displacement
or escape of the merged BH due to gravitational radiation
recoil, and to the formation of a lower-density stellar core
(Merritt \etal\ 2004).
It is possible that some late-type galaxies have
a three-tier hierarchical relation between a bulge,
a NSC, and an IMBH inside the NSC. Theoretical arguments
have suggested that if an IMBH forms via core collapse
in the core of a massive star cluster, its mass
should be $\sim$0.2\% of the host cluster mass, coincidentally
similar to the mass ratio between the bulge and its CMO.

In principle, the effective radius of a NSC may provide
clues on its dynamical processes. In \ngc, the NSC
is older and less compact ($r_e = 11.7$~pc)
than most other NSCs in late-type spirals (typical
$r_e \sim 3.5$~pc). If the cluster was formed with
such a large $r_e$, it would not have been compact
enough to form an IMBH via core collapse (Portegies~Zwart
\etal\ 2004). Alternatively, one may speculate
that the NSC was originally more compact and
has evolved to such a large radius because
of the presence of one or more BHs in its core
(Ebisuzaki \etal\ 1991; Milosavljevi\'{c} \etal\ 2002;
 Graham 2004).
Other physical processes not related to BH coalescence
may also cause a star cluster to expand. For example,
studies of clusters in the LMC show clusters
 are born compact and evolve to a range of \re\ as they age
 (Elson \etal\ 1989; Mackey \& Gilmore 2003). 
But studies also show
 this evolution is not due to differences in IMF (deGrijs \etal\ 2002),
 time-varying external tidal fields, nor differences in the 
 number of hard primordial binaries present (Wilkinson \etal\ 2003).
If no massive BH formed, the cluster could still evolve to large \re.
For the LMC clusters (none of which are thought to host a massive BH!),
 \re\ evolution 
 is likely due to differences in star formation efficiencies with
 lower efficiencies tending to expel gas before a dense stellar core 
 (with a population of massive stars) can form
 and leading eventually to cluster expansion
       (Goodwin 1997; Vine \& Bonnell 2003).
If evolution is through this mechanism, then we would expect a
 smooth distribution of cluster \re\ values that included the value
 $r_e=11.7$~pc found here for \ngc.
This does not appear to be the case (cf. Fig.~4 of B\"{o}ker \etal\ 2004)
 although the B\"{o}ker \etal\ sample is not large.

Another difference between NSCs with or without
a relatively massive BH is the role of feedback.
More powerful feedback is expected from a nuclear
IMBH or SMBH, leading to gas expulsions, perhaps
large-scale outflows, and cyclic episodes of star
formations. In this sense, the distribution
of stellar populations in a NSC provides a fossil
record of past galactic activity, and may be used
to reconstruct episodes of galactic mergers
or phases of nuclear activity. 
However, in the case
of a relatively low-mass NSC such as the one
in \ngc, we find that normal stellar winds and SNe
can provide enough energy to expel cool interstellar
gas from the cluster and quench star formation,
without the need of AGN feedback. 

Interestingly, \ngc\ is one of those rare galaxies with gas column densities
 below the critical value needed to sustain activity but has a normal 
 star formation rate (Martin \& Kennicutt 2001).
Perhaps a strong outflow from the central region has helped trigger 
 star formation at larger radii. 
Outflow may also account for some of the anomalous \hi\ discovered in the 
 halo of \ngc\ (Fraternalli \etal\ (2002, 2004).
These points will be addressed in a future paper (Yukita \etal, in preparation).

The time interval since the last star formation episode is longer 
 for the NSC in \ngc\ than is typical for clusters in late-type spirals.
Davidge \& Courteau (2002) also noted the lack of star formation activity 
 more recent than $\sim$100~Myr in the inner disk of \ngc\ and suggested
 it may be due to expulsion of potential star-forming material from
 the central regions. 
Indeed, active star formation is ongoing in several giant \hii\ regions
 (Drissen \etal\ 1999; Yukita \etal, in preparation) surrounding the 
 nucleus of \ngc\
 but the NSC itself contains perhaps the oldest collection of stars in 
 the region.

\begin{center}
\begin{tabular}{crlcr}
\multicolumn{5}{c}{{\sc TABLE 1}} \\
\multicolumn{5}{c}{{\sc X-ray Observations of NGC 2403 and its Nuclear Source Luminosity}} \\
\hline \hline
No. & \multicolumn{1}{c}{Mission/Instrument} & \multicolumn{1}{c}{Date/Identifier} &   \multicolumn{1}{c}{$L_{\rm X}$(0.5-8.0 keV)} & \multicolumn{1}{c}{$t_{exp}$} \\
 & & & (10$^{37}$ \ergl ) & (ks) \\
\hline
1 & \cxo/ACIS-S   & 2001-04-17/2014       & $<0.049$             & 36.1 \\
2 & \xmm/MOS$+$PN & 2003-04-30/0150651101 & $<2.8$               &  8.0 \\
3 & \xmm/MOS$+$PN & 2003-09-11/0150651201 & $69.8^{+11.3}_{-8.2}$ &  7.5 \\
4 & \cxo/ACIS-S   & 2004-08-23/4628       & $<0.005$             & 47.1 \\
5 & \xmm/MOS$+$PN & 2004-09-12/0164560901 & $<1.0$               & 60.0 \\
6 & \cxo/ACIS-S   & 2004-10-03/4629       & $2.9^{+0.1}_{-2.5}$  & 45.1 \\
7 & \cxo/ACIS-S   & 2004-12-22/4630       & $18.7^{+0.9}_{-7.0}$ & 50.6 \\
\hline
\end{tabular}
\end{center}

\begin{center}
\begin{tabular}{cccc}
\multicolumn{4}{c}{{\sc TABLE 2}} \\
\multicolumn{4}{c}{{\sc X-Ray Spectral Fit Parameters}} \\
\hline \hline
 & Observation 3 & Observation 6 & Observation 7 \\
\hline
Fitted energy range (keV)   &  0.3 - 10.0             &  0.5 - 2.0               & 0.5 - 3.5 \\
\hline
  \multicolumn{4}{c}{Disk Blackbody Model} \\
$n_H$ (10$^{21}$ cm$^{-2}$)$^a$ & $0.9^{+0.2}_{-0.2}$ &  $0.4^{+1.0}_{-0.0}$     & $0.4^{+0.7}_{-0.0}$  \\
$T_{\rm in}$ (keV)          & $1.11^{+0.09}_{-0.09}$  &  $0.46^{+0.06}_{-0.12}$  & $0.81^{+0.10}_{-0.13}$\\
$R_{\rm in}$ (km)$^b$       & $47.0^{+7.9}_{-6.6}$    &  $60.1^{+75.5}_{-15.1}$  & $45.7^{+20.4}_{-9.1}$\\
$M$/\msun\                  & 5.4                    &  7.2                    & 5.4         \\
$\dot{M}$ (10$^{-7}$ \msun\ yr$^{-1}$) & 1.24         &  0.05                    & 0.30          \\
$L_{\rm disk, bol}$ (10$^{37}$ \ergl ) & 85.4         &  4.52                    &   23.7        \\
$\chi^2$/DOF$^c$            & 73.8/97                 & 115.6/101                & 215.7/204     \\
\hline
  \multicolumn{4}{c}{Power Law Model}      \\
$n_H$ (10$^{21}$ cm$^{-2}$) &  $3.0^{+0.4}_{-0.4}$     & $2.2^{+2.1}_{-1.8}$     &  $2.1^{+1.1}_{-1.0}$  \\
$\Gamma$                    &  $2.22^{+0.13}_{-0.13}$  & $3.05^{+1.24}_{-1.15}$  &  $2.28^{+0.42}_{-0.40}$\\
$\chi^2$/DOF$^c$            &   91.7/97                & 116.5/101               & 218.2/204      \\
\hline
\multicolumn{4}{l}{$^a$Fixed at Galactic value for Observations 6 \& 7}\\
\multicolumn{4}{l}{$^b$Assumes $\cos \theta \sim 1$}\\
\multicolumn{4}{l}{$^c$C-statistic/Number of data bins for Observations 6 \& 7}
\end{tabular}
\end{center}

\end{document}